\documentclass[twoside,hidelinks]{article}
\usepackage{graphicx} 

\title{\centering Solar Flare Forecast: \par A Comparative Analysis of Machine Learning Algorithms for Predicting Solar Flare Classes}

\author{Julia Bringewald}
\date{February 2025}

\usepackage[a4paper, margin=3cm]{geometry}  
\graphicspath{images/}
\usepackage[labelfont=bf]{caption}
\usepackage{amsmath}
\usepackage{array}
\usepackage[table]{xcolor}
\usepackage{booktabs}
\usepackage{subcaption}
\usepackage{url}
\usepackage{doi}             
\usepackage{fancyhdr}
\usepackage{natbib}

\begin{document}    
\maketitle

\begin{center}
    {\flushleft \textit{Keywords:} Solar flares - Machine learning - Space weather forecast } 
\end{center}

\bigskip

\renewcommand{\abstractname}{\textbf{\MakeUppercase{Abstract}}}
\begin{abstract}
Solar flares are among the most powerful and dynamic events in the solar system, resulting from the sudden release of magnetic energy stored in the Sun’s atmosphere. These energetic bursts of electromagnetic radiation can release up to \(10^{32}\) erg of energy, impacting space weather and posing risks to technological infrastructure and therefore require accurate forecasting of solar flare occurrences and intensities. This study evaluates the predictive performance of three machine learning algorithms: Random Forest, k-Nearest Neighbors (KNN), and Extreme Gradient Boosting (XGBoost) for classifying solar flares into 4 categories (B, C, M, X). Using the dataset of 13 SHARP parameters, the effectiveness of the models was evaluated in binary and multiclass classification tasks. The analysis utilized 8 principal components (PC), capturing 95\% of data variance, and 100 PCs, capturing 97.5\% of variance. Our approach uniquely combines binary and multiclass classification with different levels of dimensionality reduction, an innovative methodology not previously explored in the context of solar flare prediction. Employing a 10-fold stratified cross-validation and grid search for hyperparameter tuning ensured robust model evaluation. Our findings indicate that Random Forest and XGBoost consistently demonstrate strong performance across all metrics, benefiting significantly from increased dimensionality. The insights of this study enhance future research by optimizing dimensionality reduction techniques and informing model selection for astrophysical tasks. By integrating this newly acquired knowledge into future research, more accurate space weather forecasting systems can be developed, along with a deeper understanding of solar physics. \footnote{Code available at \url{https://github.com/juliabringewald/Solar-Flare-Forecast}}.
\end{abstract}

\newpage

\section{Introduction}
Solar flares are among the most powerful and dynamic events in the solar system, resulting from the sudden release of energy of the Sun’s atmosphere \citet{Kontar_2011}. These energetic bursts of electromagnetic radiation originate from stressed magnetic fields in active regions (ARs). The impact of solar flares extends beyond the Sun, influencing space weather and posing risks to technological infrastructure such as satellite communications, power grids, and aviation systems (\citet{balan2016}; \citet{hayes2016}). Given their potential for disruption depending on their intensity, presented in \hyperref[table:Table 1]{Table~\ref{table:Table 1}}
, accurate forecasting of solar flare occurrences and intensities is essential for space weather prediction and planning satellite operations.\par 
As the closest star to Earth, the Sun plays a crucial role in shaping the space environment, with solar activity driving significant variations in space weather conditions. X-ray flux, a key indicator of solar activity, represents high-energy emissions from solar flares and other dynamic processes on the Sun’s surface. Predicting these variations is critical for safeguarding space missions, satellite operations, and ground-based technological infrastructure. Solar flares, alongside coronal mass ejections (CMEs), can trigger geomagnetic and particle disturbances, leading to technological and societal consequences (\citet{daglis2004}). Observational and theoretical studies indicate that solar flares and CMEs are powered by the rapid release of magnetic free energy stored in the corona, facilitated by magnetic reconnection (\citet{priest2002}). The build-up of this energy is caused by the evolution of the magnetic field on the photosphere. Although direct measurements of the coronal magnetic field remain challenging, observations of the photospheric magnetic field provide critical insights into energy accumulation and flare-triggering mechanisms (\citet{wang2019}). Various parameters, including sunspot classification, vertical electric currents, magnetic free energy, shear, and helicity, characterize ARs and serve as indicators for solar flare prediction. \par 
Despite significant research efforts, the connection between AR magnetic complexity and flare productivity remains only partially understood. However, the fundamental coupling between the photosphere and the corona has motivated the use of photospheric magnetic field parameters in flare prediction models, particularly through statistical and machine learning techniques (\citet{Bloomfield_2012}; \citet{barnes2016}). 
Unlike traditional statistical methods that require assumptions about data distribution, machine learning algorithms can learn the inherent patterns in data, which presents a beneficial approach for predicting space weather.
Many previous studies have relied on parameters derived from the line-of-sight (LOS) component of the photospheric magnetic field to estimate the probability of flare occurrence as well as the intensity (e.g., \citet{gallagher2002}; \citet{Bloomfield_2012}). \par
The SHARP (Space Weather HMI Active Region Patch), derived from NASA's Solar Dynamics Observatory (SDO) Helioseismic and Magnetic Imager (HMI), provide continuous high-cadence measurements of the Sun’s vector magnetic field. Vector magnetic data provides a more comprehensive view than scalar data by including the direction (inclination and declination) of the magnetic field. \par 
These data enable the computation of key magnetic parameters used in flare forecasting. This study utilized SDO/HMI data to predict the maximum magnitude of solar flares classified by GOES classes (B, C, M, X) (by \citet{liu2017dataset}). It should be emphasized that the data set utilized in this research is publicly accessible.
\newline

\begin{table}[h]
    \centering
    
    \small
    \renewcommand{\arraystretch}{1.1} 
    \begin{tabular}{|>{\centering\arraybackslash}p{2.5cm}|>{\centering\arraybackslash}p{3cm}|>{\centering\arraybackslash}p{5cm}|}
        \hline
        \textbf{Solar flare classification} & \textbf{Associated X-ray flux - I (W/m$^2$)} & \textbf{Possible effects on Earth} \\
        \hline
        B & $I < 1E{-06}$ & none \\
        \hline
        C & $1E{-06} \leq I < 1E{-05}$ & Possible effects on space missions. \\
        \hline
        M & $1E{-05} \leq I < 1E{-04}$ & Blackout in radio transmissions and possible damage to astronauts outside spacecraft. \\
        \hline
        X & $I \geq 1E{-04}$ & Damage to satellites, communication systems, power distribution stations, and electronic equipment. \\
        \hline
    \end{tabular}
    \caption{\textbf{Solar flare intensity and associated effects}}
    \label{table:Table 1}
\end{table}

We compare the predictive performance of three machine learning classification models - Random Forest, k-Nearest Neighbors (KNN), and Extreme Gradient Boosting (XGBoost) - in classifying solar flares into GOES categories. 
\par The selection of these specific models is based on their previously studied ability regarding classification tasks within the field of astrophysics. Random Forest, as indicated by \citet{cruz2024}, has been highlighted for its superior ranking capabilities. XGBoost, praised by \citet{bentejac2021}, is widely recognized for its robust performance in classification tasks. Furthermore, the K-Nearest Neighbors (KNN) classifier, as noted by \citet{ashai2022}, has consistently achieved high accuracy scores, affirming its reliability as a classification model.
\newline
\par
This study aims to introduce a new layer of comparison within the three primary machine learning models—K-Nearest Neighbors (KNN), Random Forest (RF), and Extreme Gradient Boosting (XGBoost). The performance of these models is evaluated in both binary and multiclass classification tasks. Additionally, the impact of dimensionality reduction is assessed by comparing model performance using 8 principal components (PC) versus 100 principal components derived from Principal Component Analysis (PCA), capturing different amounts of data variance.
\par Unlike previous studies, our approach is the first, to our knowledge, to compare these models across both classification types and different levels of dimensionality reduction in the context of solar flare prediction. This allows us to thoroughly examine the strengths and weaknesses of each model under different scenarios, addressing a gap that has not yet been explored in this field.
\par The findings of this study contribute to the advancement of space weather prediction, emphasizing the potential of machine learning-driven techniques to improve prediction systems for solar flares. Enhanced forecasting capabilities can mitigate the effects of extreme space weather events and improve our understanding of solar physics and contribute to better space weather models.

\section{Dataset}
\par For this study, a pre-processed dataset of solar flares based on Space-weather HMI Active Region Patches (SHARP) data, originally created by \citet{liu2017dataset}, was utilized. The SHARP dataset, developed by the SDO/HMI team (\citet{Bobra_2015}), provides magnetic field measurements and derived parameters for automatically identified and tracked active regions (ARs). Additionally, the cgem. Lorentz data series provides estimations of Lorentz forces to help analyzing AR dynamics (\citet{fisher2012}).
\par \citet{liu2017} constructed their dataset by selecting ARs that produced flares over a ~6.5-year period (May 2010–December 2016), covering the peak of Solar Cycle 24. Flare records were obtained from the GOES X-ray flare catalogs provided by the National Centers for Environmental Information (NCEI). These records were merged with USAF Solar Observing Optical Network H-alpha flare listings (\citet{denig2012}), ensuring comprehensive flare identification and classification.
\par The dataset follows the GOES flare classification (B, C, M, X), where each AR is categorized by the highest GOES-class flare it produced (\citet{song2009}; \citet{yuan2010}). To ensure data quality, only C-, M-, and X-class ARs with flares occurring within ±70° of the central meridian were included, while B-class ARs were manually verified through solarmonitor.org. Furthermore, the dataset includes 13 key magnetic field parameters, presented in \hyperref[table:Table 2]{Table~\ref{table:Table 2}}
, identified by \citet{Bobra_2015} as strong predictors of solar flare activity.

\renewcommand{\arraystretch}{1.1} 
\begin{table}[h]
    \centering
    \small
    \begin{tabular}{|l|p{8.5cm}|}
        \hline
        \rowcolor{gray!35}
        \textbf{PARAMETER} & \textbf{DESCRIPTION} \\
        \hline
        ABSNJZH & {Absolute value} of net current helicity \\\hline\rowcolor{gray!15}
        AREA\_ACR & Area of strong field pixels in active region (AR) \\\hline
        EPSZ &Sum of z-component of normalized Lorentz force \\ \hline\rowcolor{gray!15}
        MEANPOT & Mean photospheric magnetic free energy \\\hline
        R\_VALUE & Sum of flux near polarity inversion line \\\hline\rowcolor{gray!15}
        SAVNCPP & Sum of the modulus of the net current per polarity \\\hline
        SHRGT45 & Fraction of area with shear \( >45^\circ \) \\\hline\rowcolor{gray!15}
        TOTBSQ & Total magnitude of Lorentz force \\\hline
        TOTFZ & Sum of z-component of Lorentz force \\\hline\rowcolor{gray!15}
        TOTFOT & Total photospheric magnetic free energy density \\\hline
        TOTUSJH & Total unsigned current helicity \\\hline\rowcolor{gray!15}
        TOTUSJZ & Total unsigned vertical current \\\hline
        USFLUX & Total unsigned flux \\
        \hline
    \end{tabular}
    \caption{Overview of SDO/HMI Magnetic Parameters  }
    \label{table:Table 2}
\end{table}

\par By utilizing this well-validated dataset, the study focuses on applying machine learning techniques to improve solar flare prediction, evaluating classification models based on their ability to forecast flare intensity.

\par 
\section{Methodology}

\subsection{Experimental Setup}
\par The dataset consists of solar activity parameters, including sunspot counts, magnetic field measurements, and previous flare occurrences. Several machine learning models, including logistic regression, decision trees, and neural networks, are trained and evaluated based on their predictive accuracy.

\subsubsection{Feature engineering}
\par
\begin{figure}[htbp]
    \centering
    \includegraphics[width=10cm]{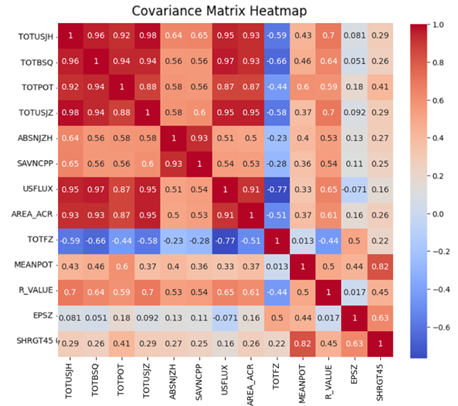}
    \caption{Covariance heatmap of magnetic parameters.
\textit{Note:} A covariance heatmap is a visual representation of the covariance matrix of a dataset's features. Covariance measures the strength of joint variability between two or more variables, as to say how much two variables change together. Positive covariance indicated that as one variable increases, the other tends to increase as well. Negative covariance indicates that as one variable increases, the other tends to decrease.}
    \label{fig:Figure 1}
\end{figure}

 PCA (Principal Component Analysis) method is used for reduction of dimensionality of the parameter space or as to say the number of variables. The 13 magnetic parameters are standardized using StandardScaler, because PCA relies on the data being normally distributed with each feature having a similar range. After standardizing the data, interaction features of the standardized data are calculated, visualized in 
\hyperref[fig:Figure 1]{Figure~\ref{fig:Figure 1}}. Interaction features can help capture non-linear relationships between the original features that may not be apparent when considering them individually. Also, this approach helps to reduce overfitting by capturing the underlying structure of the data more effectively.

How many of the interaction features are most beneficial for the model quality, is determined by a certain variance threshold of the data the features should capture. In the context of machine learning variance refers to how much the feature values vary or are spread out from their mean. Capturing as much variance as possible refers to preserving the diversity of the data. This is important to keep the full representation of data diversity while reducing your initial amount of features (in this case the interaction features). In simple terms, capturing a lot of variance in your dataset is like summarizing a book while keeping all the main ideas and discarding the unnecessary details.

\par
\begin{figure}[htbp]
    \centering
    \includegraphics[width=10cm]{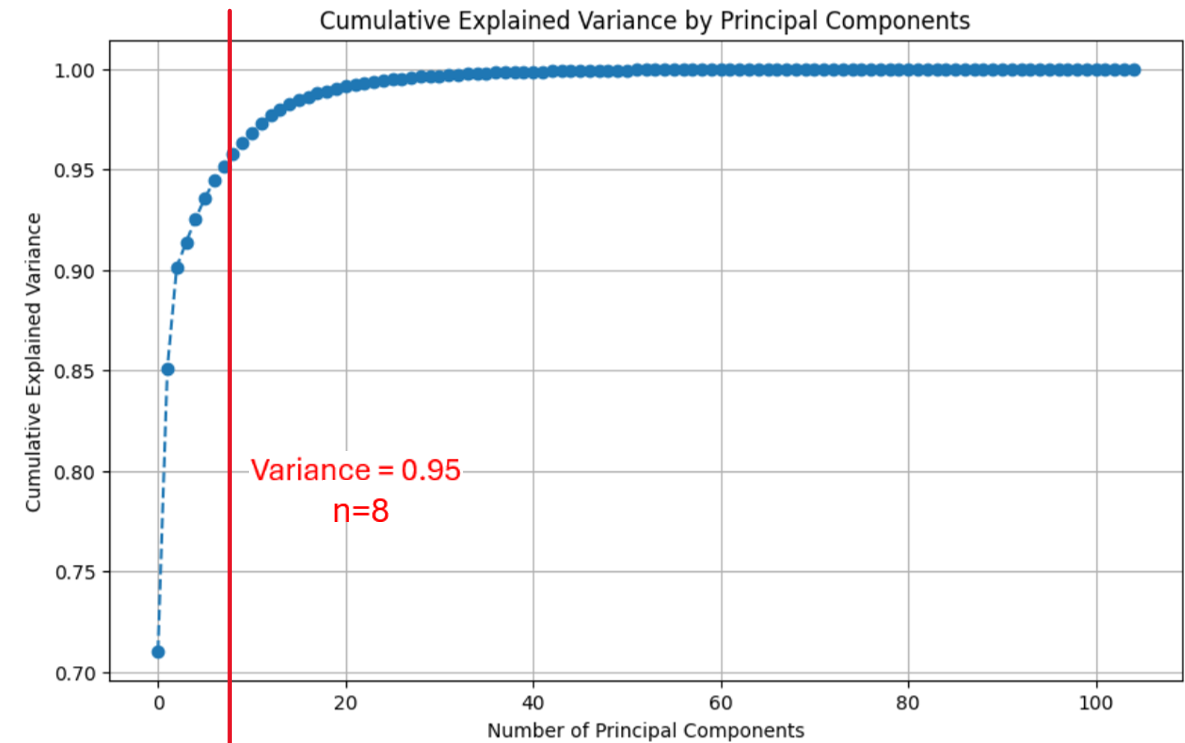}
    \caption{Cumulative Explained Variance by Principal Components (Interaction features)}
    \label{fig:Figure 2}
\end{figure}

 In relation to covariance calculation, illustrated in \hyperref[fig:Figure 2]{Figure~\ref{fig:Figure 2}}, PCA derives the features with the strongest covariance of interaction features, while capturing a determined amount of variance from the original data.
\par
\begin{figure}[htbp]
    \centering
    \includegraphics[width=16cm]{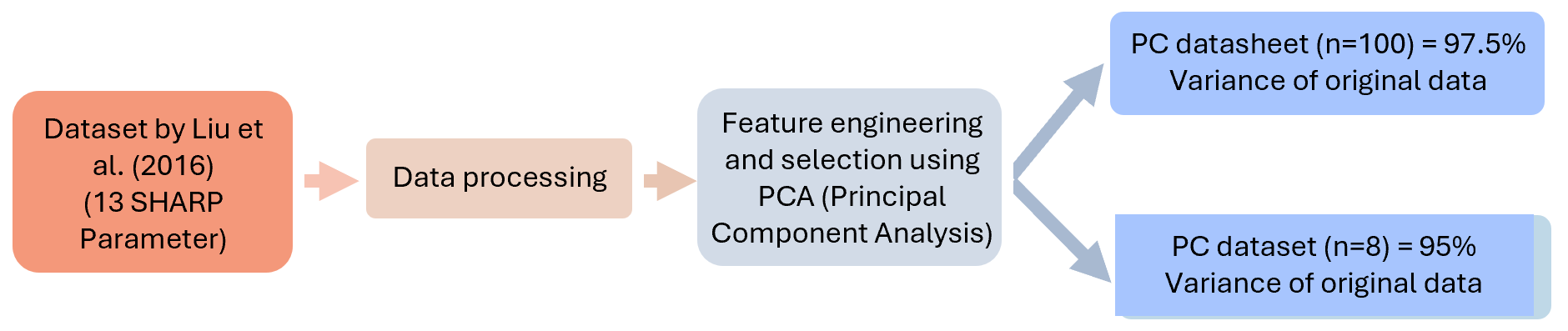}
    \caption{Visualization of the feature engineering process}
    \label{fig:Figure 3}
\end{figure}

 In this study, a threshold of 95\% was used, resulting in a total number of n=8 components necessary to achieve this, as \hyperref[fig:Figure 2]{Figure~\ref{fig:Figure 2}} illustrates with the cumulative explained variance. This refers to the total amount of information or variability captured by a certain number of principal components in the interaction feature dataset. To observe the impact of capturing more variance, and therefore more original information from the original dataset, the use of 100 components was analyzed, capturing 97.5\% of the variance. This process is visualized in \hyperref[fig:Figure 3]{Figure~\ref{fig:Figure 3}} The analysis did not include more than 100 components due to limitations in computational resources.

\samepage

This approach examines how using different amounts of original data and the risk of overfitting from too detailed data affect model performance. 
These principal components (PC=8 and PC=100) are selected and retained in two different data sets, which will be further used in the machine learning process.

\subsection{Algorithms}
\subsubsection{K-Nearest Neighbours (KNN)}

\par The K-Nearest Neighbours (KNN) algorithm is a supervised learning classifier, introduced by Cover and Hart in 1967. It operates on the principle of proximity, predicting the output based on the most similar data points in the feature space. For classification, KNN assigns a class label to a new sample based on the majority vote among its K nearest neighbours, where K represents the number of neighbours considered. This can be more accurately described as “plurality voting” in multi-class settings.
The algorithm's simplicity and intuitive approach make it popular and widely applied across various fields. KNN requires recalculating the distance between the input and all data points whenever new data is introduced, which can be computationally intensive compared to other models.
\par
Key hyperparameter of KNN include the number of nearest neighbours (K), the weighting of each neighbour (whether all neighbours contribute equally or closer neighbours have a larger weight), and the distance metric used for calculating proximity between samples.

\subsubsection{Random Forest classifier}
The Random Forest (RF) algorithm, trademarked by Leo Breiman and Adele Cutler, is a widely used machine learning method known for its flexibility and ease of use. It is an ensemble learning technique that combines the output of multiple decision trees to achieve a single result, making it suitable for both classification and regression tasks.
\par In the training phase, each decision tree in the forest is trained on a bootstrap sample of the original data, where approximately 37\% of instances are duplicated. Additionally, at each node split, a random subset of the input features is selected to determine the best split. This dual randomization—bootstrap sampling and feature selection—helps reduce overfitting and increase the model's robustness.
For classification tasks, the final prediction is determined by majority voting among the trees.

\subsubsection{Extreme Gradient Boosting (XGBoost)}
XGBoost (eXtreme Gradient Boosting) is a distributed, open-source machine learning library that excels in speed, efficiency, and scalability for large datasets. It employs gradient boosted decision trees, a supervised learning algorithm utilizing gradient descent for optimization. XGBoost starts with a weak learner, known as the base learner, and builds new trees in an additive manner to correct the base learner’s errors. The algorithm calculates the residuals, which are the differences between the predicted and actual values, and uses these residuals to guide the creation of subsequent trees. The model's performance is evaluated using loss functions, such as mean squared error for regression and cross-entropy loss for classification. Gradient descent minimizes the loss, improving model performance with each iteration. XGBoost's ability to handle both classification and regression tasks, combined with its efficient handling of large datasets and robust performance, makes it a popular choice in machine learning applications. \newline
\par In simpler words: XGBoost begins with a simple model that makes basic predictions. Then, it adds new models (decision trees) in a step-by-step process, each one learning from the mistakes of the previous models. These trees try to fix the errors made by earlier predictions. The predictions from all the trees are combined to make a final, more accurate prediction. The algorithm uses a technique called gradient descent to adjust and minimize the errors, making the model better with each new tree added.

\subsection{Training and Evaluation}

\par
\begin{figure}[h]
\begin{flushright}
    \includegraphics[width=12cm,height=10cm]{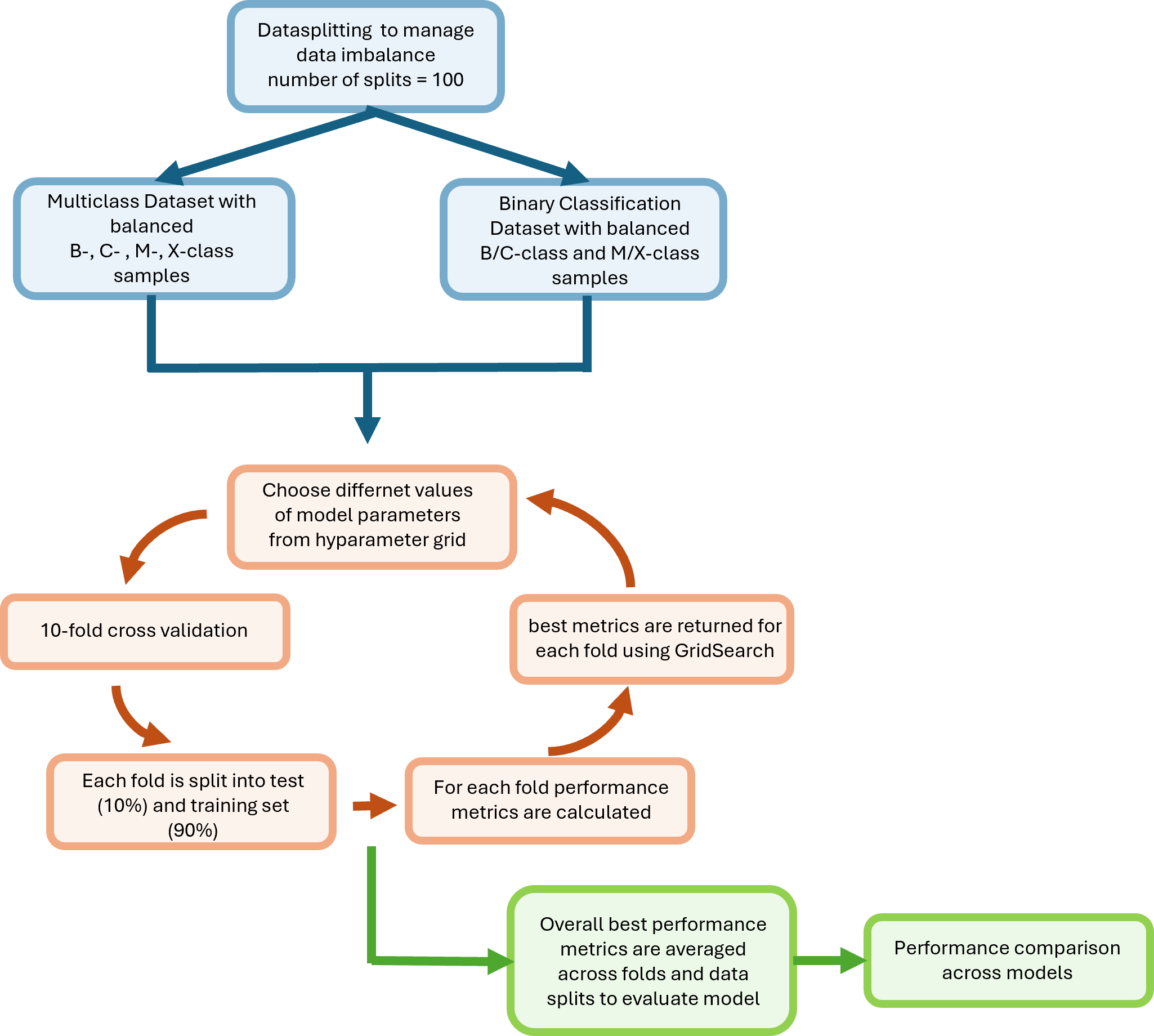}
    \caption{Visualization of models' training an evaluation method}
    \label{fig:Figure 7}
\end{flushright}
\end{figure}

 For the flare prediction, every algorithm operates in classification mode. During the training phase, each input training sample includes interaction parameter values (n=8 or n=100 as described in 3.1. Experimental Setup) of an AR and its corresponding maximum GOES flare class, which represents the target variable. In the testing/validation phase, these parameters of an AR sample serve as the model input, and the classifier predicts the GOES class of this AR.
The dataset utilized, created by \citet{liu2017dataset}, exhibits an imbalance in the number of AR samples, with 552 C-class samples, 23 X-class samples, 142 M-class samples, and 128 B-class samples. To address this class imbalance, which poses a significant challenge in machine learning (e.g., \citet{japkowicz2002}), the approach suggested by \citet{liu2017} was adopted. From the total 552 C-class samples, 142 unique samples were randomly selected to create a balanced dataset. To ensure robustness and avoid bias, this random selection was repeated 100 times, producing 100 datasets for multi-class classification. Each dataset contains 128 B-class, 142 C-class, 142 M-class, and 23 X-class AR samples, as illustrated in the models' training method in \hyperref[fig:Figure 7]{Figure~\ref{fig:Figure 7}}.

\par

\par Additionally, for binary classification, another set of 100 datasets was constructed to facilitate comparisons with previous work. The B- and C-class samples were combined to form the B/C class, while the M- and X-class samples were combined to form the M/X class. For each of the 100 datasets, 165 unique B/C-class AR samples were randomly selected and paired with 165 M/X-class samples, ensuring balanced binary classification datasets. This data splitting approach prevents misclassifications by machine learning models due to the previous class imbalance.
\par Given that M/X-class solar flares produce the most significant effects on Earth, as presented in \hyperref[table:Table 1]{Table~\ref{table:Table 1}}, adopting a binary classification approach is beneficial for predicting the most intense solar flares. This method improves the prediction of solar flares that have substantial terrestrial impacts.

\par To evaluate the performance of various machine learning algorithms, the widely used 10-fold cross-validation (CV) method is applied. For each dataset, stratified 10-fold partitioning is performed using StratifiedKFold(), ensuring that each fold contains nearly equal-sized groups with balanced class distributions. The algorithm is then trained using nine folds of the data, while the remaining fold is used for validation. This process is repeated for each combination of hyperparameters being evaluated using the GridSearch() function. Hyperparameters are settings that control the learning process and are not learned from the data. Proper tuning helps improve model accuracy, reduce overfitting, and ensure that the model generalizes well to new and unseen data.
\par During the cross-validation, the model's performance is assessed using predefined metrics such as accuracy, ROC AUC, F1 score, and precision-recall AUC . The indepth functioning of the performance metrics is described as following. 
\newline
\par 
\vspace{10pt}

\begin{center}
    \begin{tabular}{rl}
       
        \textbf{Overall Accuracy} & \( = \frac{\text{TP} + \text{TN}}{\text{TN} + \text{FP} + \text{FN} + \text{TP}} \) \\[10pt]
        
        \textbf{F1 score} & \( = \frac{2 \times \text{TP}}{2 \times \text{TP} + \text{FP} + \text{FN}} \) \\[10pt]
        \\{Precision} & \( = \frac{\text{TP}}{\text{TP} + \text{FP}} \) \\[10pt]
        {Recall} & \( = \frac{\text{TP}}{\text{TP} + \text{FN}} \) \\[10pt]
        
        \textbf{PR AUC} & \( = \int_{0}^{1} \text{Precision} \, d(\text{Recall}) \)\\
        
        \\\newline\\{True Positive Rate (TPR)} & \( = \frac{\text{TP}}{\text{TP} + \text{FN}} \) \\[10pt]
        {False Positive Rate (FPR)} & \( = \frac{\text{FP}}{\text{FP} + \text{TN}} \) \\[10pt]
        \textbf{ROC AUC} & \( = \int_{0}^{1} \text{TPR} \, d(\text{FPR}) \)
       
    \end{tabular}

\begin{enumerate}
    \item True positives (TP): number of samples correctly predicted as “positive”.
    \item False positives (FP): number of samples wrongly predicted as “positive”.
    \item True negatives (TN): number of samples correctly predicted as “negative”.
    \item False negatives (FN): number of samples wrongly predicted as “negative”.
\end{enumerate}
\end{center}

\vspace{10pt}

\par The performance metrics from the 10 folds are averaged to obtain the mean performance metric for each hyperparameter combination. This averaging process mitigates the impact of data variability and provides a robust evaluation.
\par 
The hyperparameter combination that yields the highest mean performance
metric across the 10 folds is selected as the best-performing set of hyperparameters.
Various standard performance metrics, such as Receiver Operator Curve (ROC AUC), accuracy, and precision-
recall AUC and F1 are calculated to assess model performance, AUC signifying Area Under the Curve.
with values closer to 1
indicating better performance.

\newpage
\subsubsection{Hyperparameters}

Three algorithms were invested in this study, and the details about the hyperparameter search grid being listed in \hyperref[table:Table 10]{Table~\ref{table:Table 10}}. 

\renewcommand{\arraystretch}{1.1} 

\begin{table}[h]
    \centering
    \small
    \setlength{\tabcolsep}{8.5pt} 
    \rowcolors{3}{white!95!gray!20}{white} 
    \begin{tabular}{ |l|l|l| }
        \hline
        \rowcolor{gray!30} \textbf{Algorithm} & \textbf{Tuning Parameter} & \textbf{Search Grid} \\ 
        \hline
        {KNN} & Number of Nearest Neighbours (n\_neighbors) & [3,5,7,9] \\ \cline{2-3}
        & Metric of weights & [‘uniform’, ’distance’] \\ \cline{2-3}
        & Metric of distance & [‘euclidean’, ‘manhattan’] \\ 
        \hline
        {Random Forest} & Number of features & [8,9,10,11,12,13] \\ \cline{2-3}
        & Depth of trees (min\_samples\_split) & [8, 10] \\ 
        \hline
        {XGBOOST} & Learning rate & [0.1, 0.9] \\ \cline{2-3}
        & Subsample & [0.01, 1] \\ 
        \hline
    \end{tabular}
    \caption{Algorithm Hyperparameter Search Grid}
    \label{table:Table 10}
\end{table}

\subparagraph{\textit{Notes on Hyperparameters per model:}}\mbox{}\\

\textit{K-Nearest Neighbors (KNN)}
\begin{itemize}
    \item Number of Nearest Neighbours : This parameter determines how many neighbours are considered when making a prediction. The choice of the number of neighbors directly affects the model's bias-variance trade-off. A smaller number of neighbors (e.g., 3) can result in a model that captures local patterns well but might be sensitive to noise (low bias, high variance). On the other hand, a larger number of neighbors (e.g., 10) can smooth out predictions, making the model more robust to noise but potentially less sensitive to local variations (high bias, low variance).
    \item Metric of weights: This parameter decides how the distance between points affects their influence on the prediction. Using 'uniform' weights means all neighbors contribute equally, which is suitable when the data is uniformly distributed. 'Distance' weighting allows closer neighbors to have a greater influence, which can be beneficial when the relevance of neighbors decreases with distance.
    \item Metric of distance: This parameter defines how the distance between points is calculated. 'Euclidean' distance is the most commonly used metric and is appropriate when the feature space is continuous. 'Manhattan' distance is useful when dealing with high-dimensional data or when features are on different scales.
\end{itemize}

\mbox{}\\

\textit{Random Forest}
\begin{itemize}
    \item Number of features: This parameter specifies how many features are considered when splitting a node. More features can lead to a more complex model, while fewer features can simplify it.
    \item Depth of trees: This parameter controls the maximum depth of each tree in the forest. Deeper trees can capture more complex patterns but may also overfit the data. A depth of 10 for example allows the model to capture complex patterns without becoming too complex and overfitting.
    \item Minimum number of sample splits: This parameter determines the minimum number of samples required to split an internal node. Setting this parameter to a higher value (e.g., 5) ensures that splits are only made when there is enough data to justify the split, preventing overfitting and improving the model's generalization ability.
\end{itemize}

\mbox{}\\

\textit{XGBoost}
\begin{itemize}
    \item Learning rate: This parameter controls the step size at each iteration while moving toward a minimum of the loss function. A smaller learning rate can lead to more accurate models but requires more iterations.
    \item Subsample: This parameter specifies the fraction of samples used for fitting individual base learners. Lower values can prevent overfitting by introducing randomness.
\end{itemize}
Because of the considerable computational demands, only a restricted set of hyperparameters were investigated in the process of this study. Generally, all of the models (KNN, RF, XGBOOST) are trained via scikit-learn (\citet{pedregosa2011}).

\section{Results}
After the described training and evaluation process described in 3.3 Training and
Evaluation Strategy. \hyperref[table:Table 4]{Table~\ref{table:Table 4}} is the summary of these efforts.

\begin{table}[h]
    \centering

    \setlength{\tabcolsep}{8.5pt}
    \rowcolors{3}{gray!15}{white}
    \begin{tabular}{|l|c|c|c|c|c|c|}
        \hline
        \rowcolor{gray!35} \textbf{Performance Metric} & \multicolumn{2}{c|}{\textbf{Random Forest}} & \multicolumn{2}{c|}{\textbf{KNN}} & \multicolumn{2}{c|}{\textbf{XGBOOST}} \\
        \hline
        \rowcolor{gray!35} & \textbf{8 PC} & \textbf{100 PC} & \textbf{8 PC} & \textbf{100 PC} & \textbf{8 PC} & \textbf{100 PC} \\
        \hline
        \multicolumn{7}{|l|}{\textbf{Multiclass Classification:}} \\
        \hline
        Accuracy & 0.541 & 0.623 & 0.561 & 0.638 & 0.570 & 0.624 \\
        ROC AUC & 0.855 & 0.839 & 0.754 & 0.817 & 0.770 & 0.846 \\
        PR AUC & 0.557 & 0.668 & 0.567 & 0.647 & 0.562 & 0.673 \\
        F1 Score & 0.530 & 0.610 & 0.538 & 0.603 & 0.554 & 0.606 \\
        \hline
        \multicolumn{7}{|l|}{\textbf{Binary Classification:}} \\
        \hline
        Accuracy & 0.679 & 0.738 & 0.680 & 0.594 & 0.690 & 0.733 \\
        ROC AUC & 0.743 & 0.804 & 0.735 & 0.625 & 0.758 & 0.811 \\
        PR AUC & 0.790 & 0.790 & 0.758 & 0.636 & 0.787 & 0.834 \\
        F1 Score & 0.671 & 0.735 & 0.640 & 0.373 & 0.680 & 0.723 \\
        \hline
    \end{tabular}
    \caption{Model performance for binary and multiclass classification with n=8 and n=100 PC}
    \label{table:Table 4}
    
\end{table}

\subsection{Model Performance with varying Principal Components}
The performance metrics of the machine learning models — Random Forest, KNN, and XGBoost — were evaluated for both binary and multiclass classification tasks using 8 principal components (PC) and 100 principal components (PC) (\hyperref[table:Table 4]{Table~\ref{table:Table 4}}). The models exhibited notable differences in performance across these configurations. 
\par This analysis highlights the significant impact of using different numbers of interaction features from principal component analysis (PCA) on the performance of various machine learning models. By comparing algorithms such as Random Forest, KNN, and XGBoost using 8 and 100 principal components, variations in model performance can be observed based on different levels of feature complexity. While Random Forest and XGBoost demonstrate enhanced performance with 100 principal components (PC) compared to 8 PC, showing improvements in accuracy, ROC AUC, and F1 scores, KNN exhibits a different pattern. For multiclass classification, KNN shows modest improvement with increased principal components. However, for binary classification, KNN's performance declines across all metrics when using 100 PC compared to 8 PC.
\par
\begin{figure}[htbp]
    \centering
    \begin{subfigure}{0.49\textwidth}
        \centering
        \includegraphics[width=8cm]{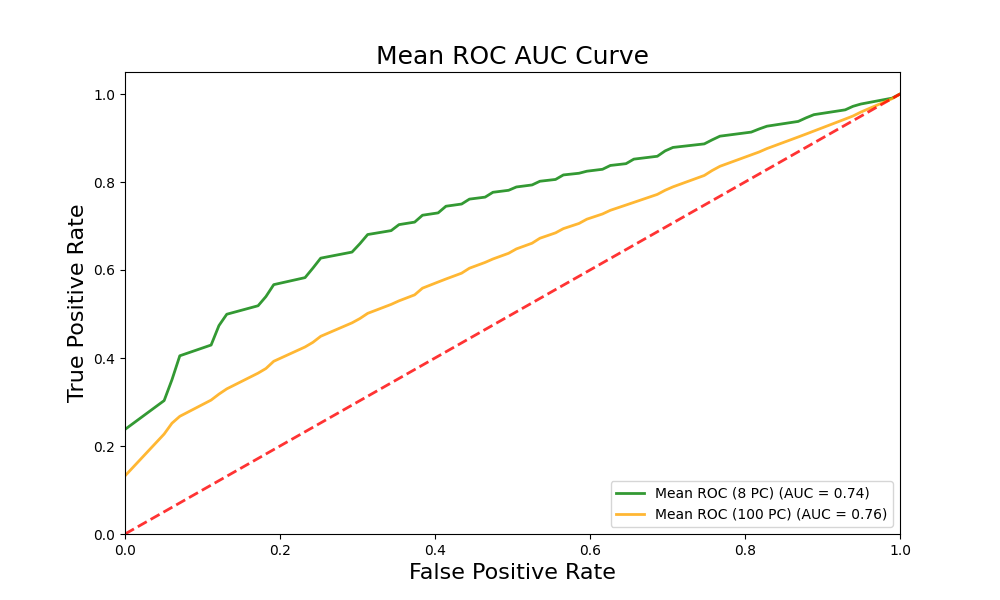}
        \caption{ROC AUC Curve}
        \label{fig:first}
    \end{subfigure}
    \hfill
    \begin{subfigure}{0.49\textwidth} 
        \centering
        \includegraphics[width=8cm]{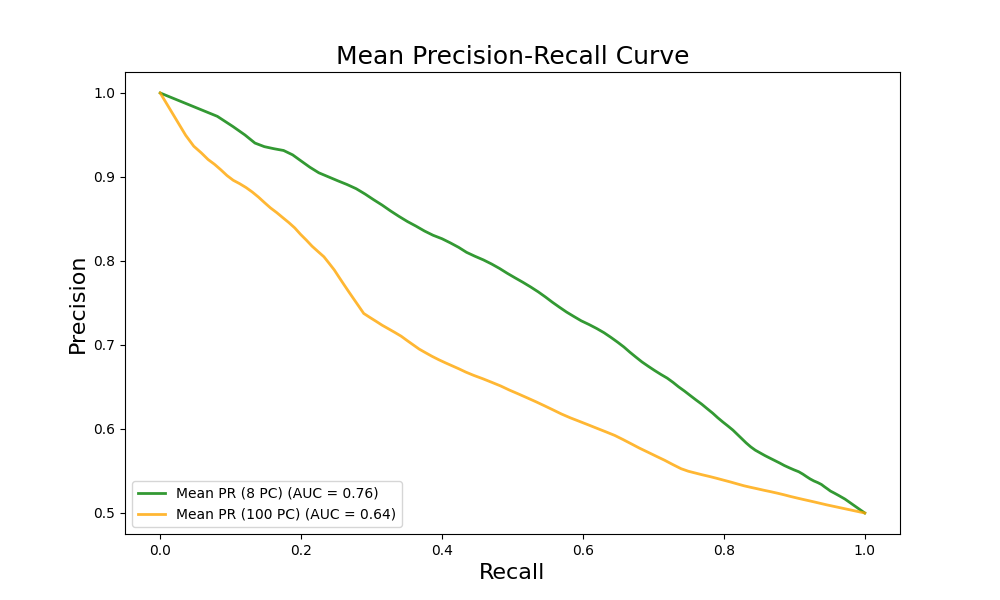}
        \caption{Precision-Recall Curve }
        \label{fig:second}
    \end{subfigure}
    \caption{Performance Curves KNN}
    \label{fig:Figure 8}
\end{figure}

\begin{figure}[htbp]
    \flushleft
    \begin{subfigure}{0.49\textwidth}
        \centering
        \includegraphics[width=8cm]{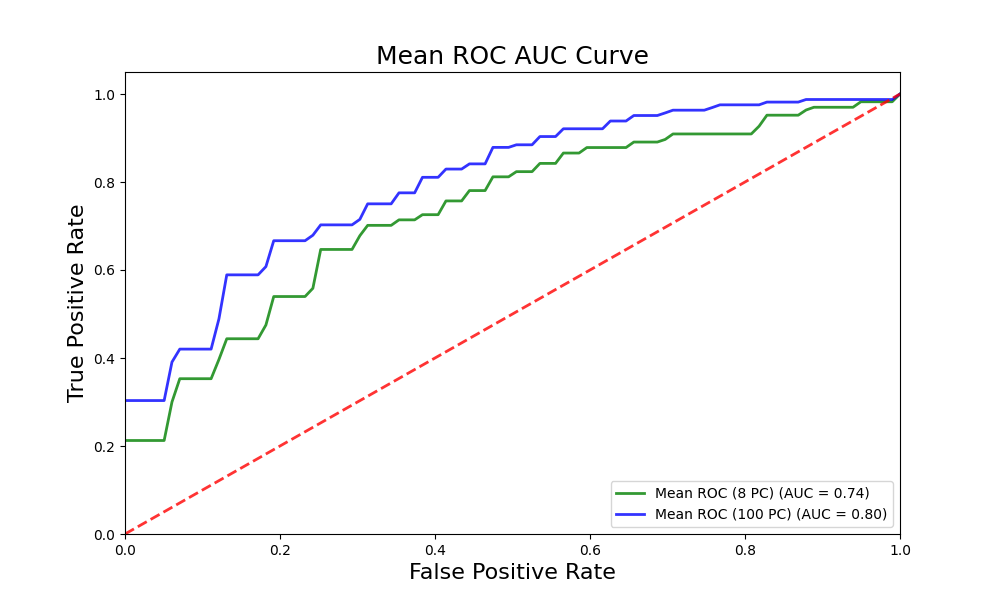}
        \caption{ROC AUC Curve}
        \label{fig:third}
    \end{subfigure}
    \hfill
    \begin{subfigure}{0.49\textwidth} 
        \centering

        \includegraphics[width=8cm]{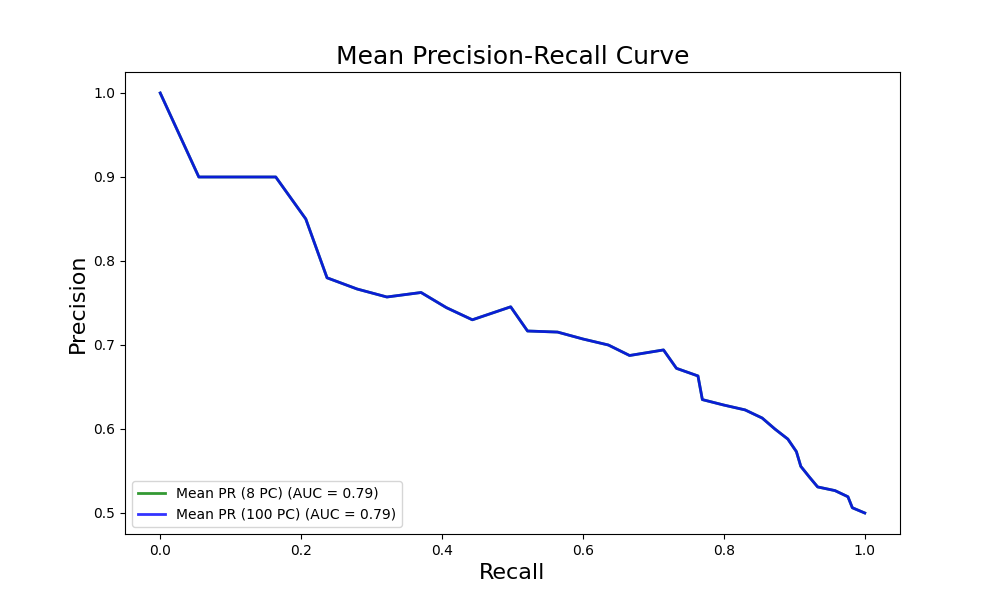}
        \caption{Precision-Recall Curve }
        \label{fig:fourth}
        
    \end{subfigure}
    
    \caption{Performance Curves Random Forest}
    \label{fig:Figure 9}
\end{figure}

\begin{figure}[htbp]
    \flushleft
    \begin{subfigure}{0.49\textwidth}
        \centering
        \includegraphics[width=8cm]{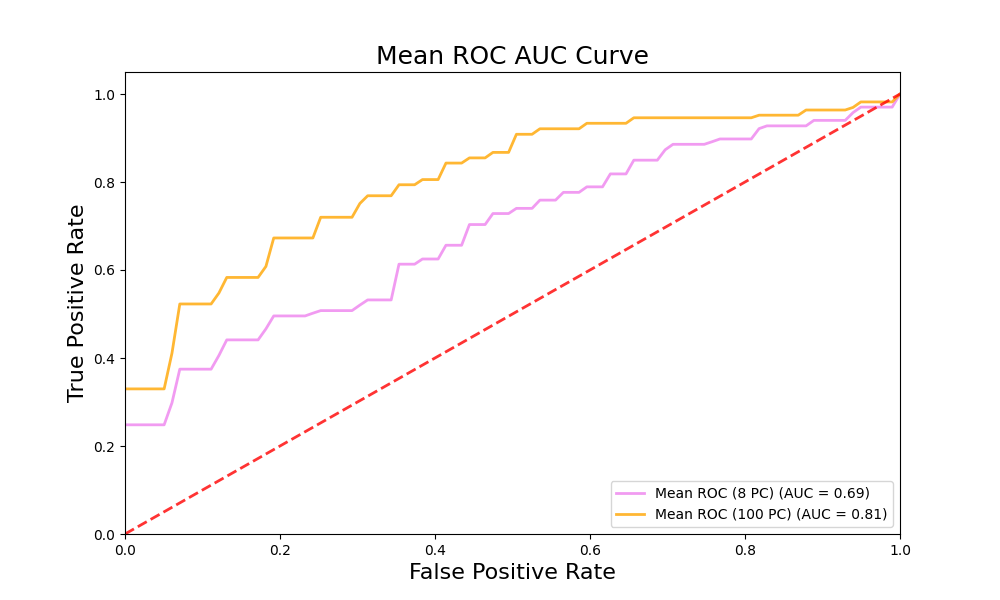}
        \caption{ROC AUC Curve}
        \label{fig:fifth}
    \end{subfigure}
    \hfill
    \begin{subfigure}{0.49\textwidth} 
        \centering

        \includegraphics[width=8cm]{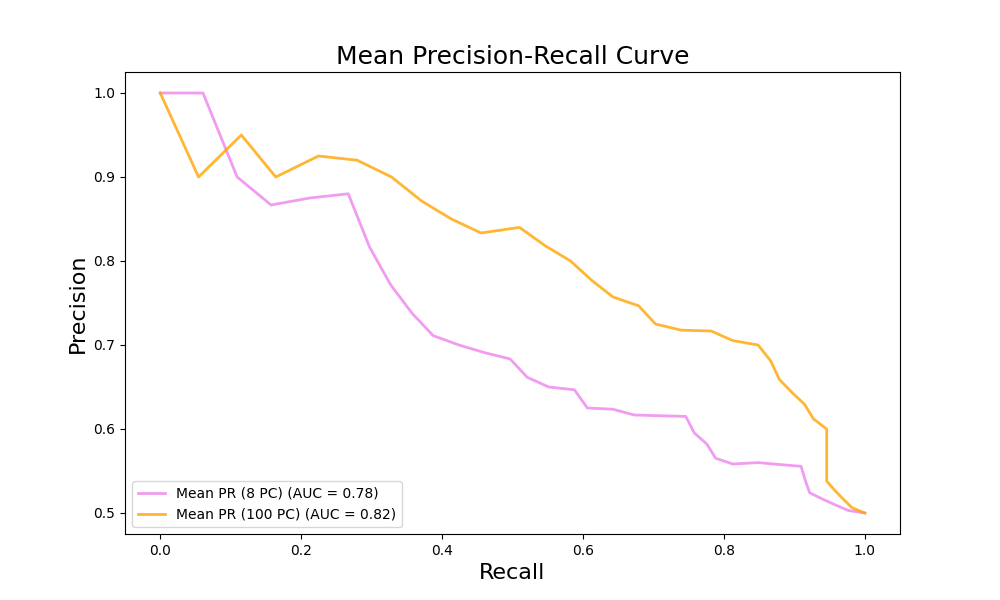}
        \caption{Precision-Recall Curve }
        \label{fig:sixth}
        
    \end{subfigure}
    
    \caption{Performance Curves XGBoost}
    \label{fig:Figure 10}
\end{figure}

 A visual representation of each model’s ability (ROC AUC and Precision-Recall Curve) to discriminate between classes (binary classification) with different feature sets and  therefore increased dimensionality, are shown in \hyperref[fig:Figure 8]{Figure~\ref{fig:Figure 8}}, \hyperref[fig:Figure 9]{Figure~\ref{fig:Figure 9}} and \hyperref[fig:Figure 10]{Figure~\ref{fig:Figure 10}}.

\subsection{Model Performance with Varying Classification Tasks (PC=8)}

Additionally, the performance of the Random Forest, KNN, and XGBoost models was evaluated for both binary and multiclass classification tasks, using 8 principal components (PC) that capture the base threshold of 95\% of the variance in the data.
\par In \textit{binary classification}, with 8 principal components, the Random Forest and
XGBoost models demonstrated strong performance across metrics such as accuracy,
ROC AUC, and F1 score. These models effectively captured the essential information
from the reduced feature set, allowing them to make accurate predictions. On the other
hand, the KNN model also performed reasonably well but showed slightly lower metrics
compared to Random Forest and XGBoost, indicating a modest ability to leverage the
reduced dimensionality for binary classification.
\par For \textit{multiclass classification}, the performance of the models with 8 principal
components varied. The Random Forest and XGBoost models showed competitive
performance, maintaining relatively high accuracy and F1 scores. These models were
able to handle the complexity of the multiclass problem effectively, even with the
reduced feature set. The KNN model exhibited moderate improvement in multiclass
classification, but its performance metrics were lower than those of Random Forest and
XGBoost, suggesting that KNN had more difficulty managing the reduced feature space
for multiclass tasks.
\newline
\par Overall, with 8 principal components, Random Forest and XGBoost consistently
outperformed KNN in both binary and multiclass classification tasks.

\begin{figure}[h]
    \centering
    \includegraphics[width=15cm]{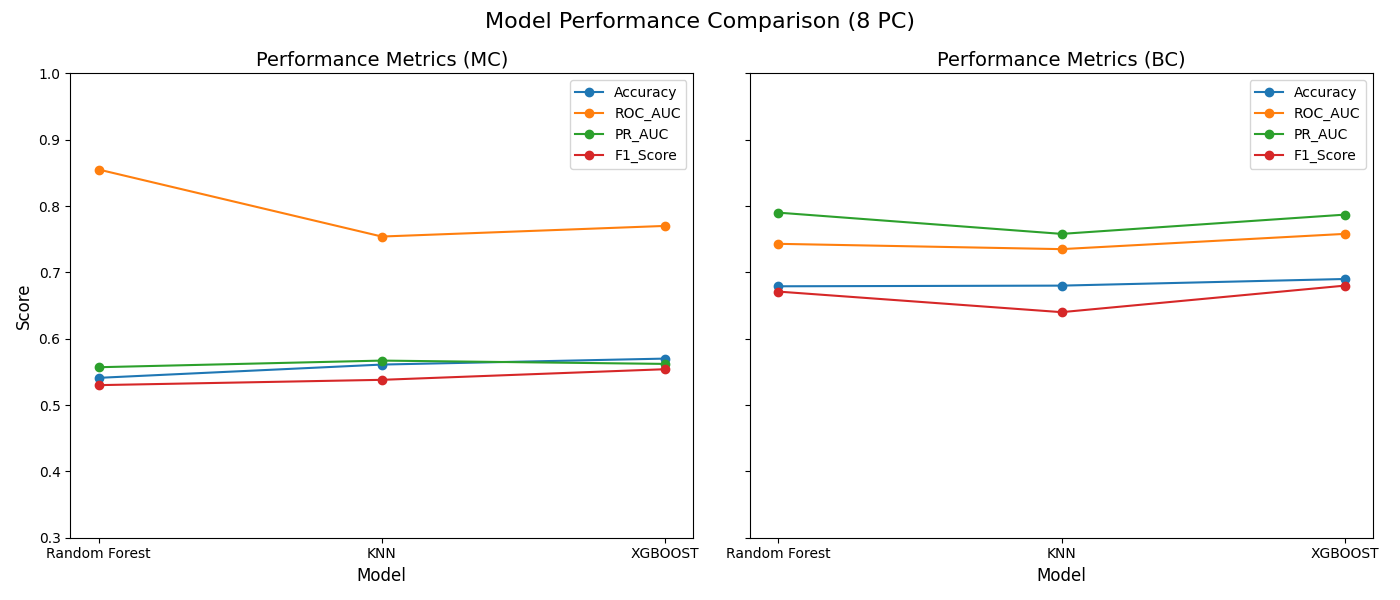}
    \caption{Model performance Comparison across varying classifications (PC=8) \textit{Note}: MC: Multiclass Classification ; BC: Binary Classification }
    \label{fig:Figure 11}
\end{figure}

\section{Discussions}
The results from \autoref{table:Table 4} regarding \textit{varying principal components}, suggest
that Random Forest and XGBoost effectively leverage the additional features provided by
100 PC (capturing 97.5\% of the variance of the original data set) to capture complex
patterns and interactions in the data, resulting in better overall performance compared
to the 8 PC approach (capturing 95\% of data set variance). The improvements in
accuracy, ROC AUC, and F1 scores indicate that these models benefit from the
increased dimensionality, allowing them to make more accurate predictions.
\par
\begin{figure}[h]
    \centering
    \includegraphics[width=15cm]{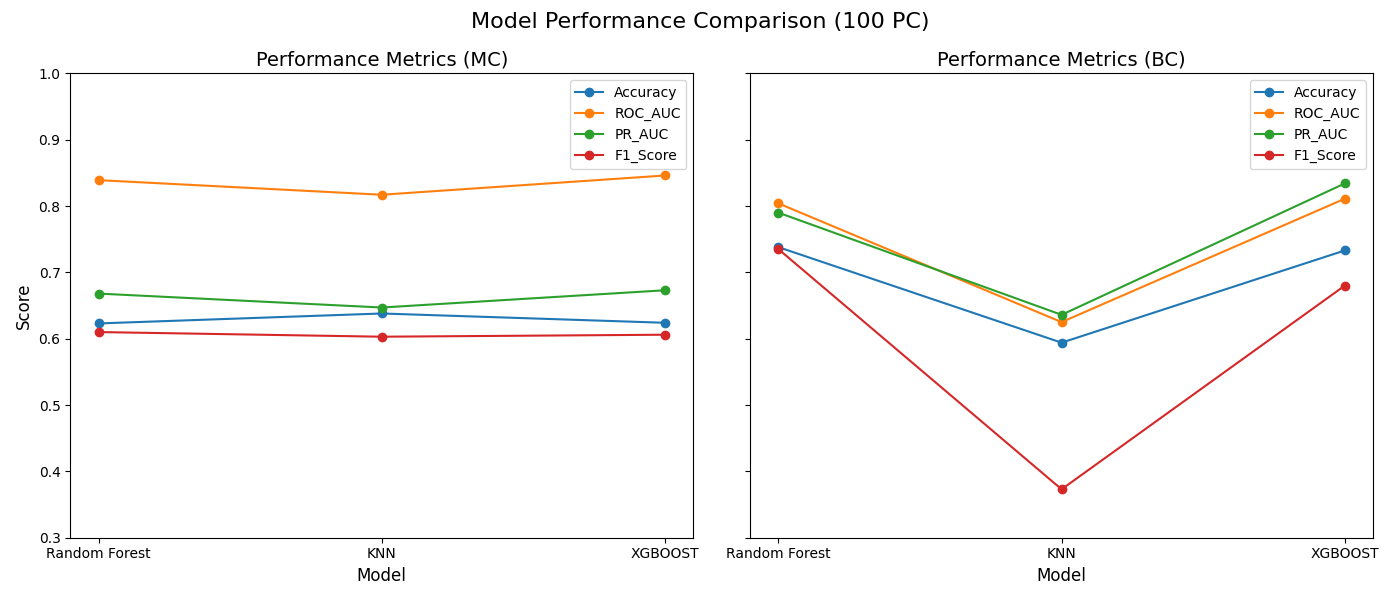}
    \caption{Model performance Comparison across varying classifications (PC=100) \textit{Note}: MC: Multiclass Classification ; BC: Binary Classification }
    \label{fig:Figure 12}
\end{figure}

 In contrast, KNN's performance behaviour indicates a sensitivity to the increased
number of features, highlghted in \hyperref[fig:Figure 11]{Figure~\ref{fig:Figure 11}} and \hyperref[fig:Figure 12]{Figure~\ref{fig:Figure 12}}. While it shows some improvement in handling multiclass
classification with more principal components, the decline in binary classification
performance suggests that KNN may struggle with the higher dimensionality, potentially
due to overfitting or difficulty in managing the increased complexity of the data.
\par These findings underscore the importance of carefully selecting the number of
interaction features derived from PCA to optimize each model's performance.

\par The analysis of \textit{varying classification (binary and multiclass)} of solar flare classes
across models highlights key insights into the performance of Random Forest, KNN, and
XGBoost models using 8 principal components (PC) for both binary and multiclass
classification tasks, visualized in \hyperref[fig:Figure 11]{Figure~\ref{fig:Figure 11}}. Random Forest and XGBoost exhibited solid performance in both
classification tasks, maintaining high accuracy, ROC AUC, and F1 scores, indicating
their ability to effectively leverage the essential features captured by the principal
components. \par In contrast, the KNN model showed moderate performance, with
reasonable results in binary classification but slightly lower metrics compared to
Random Forest and XGBoost. In multiclass classification, KNN exhibited modest
improvement but still lagged behind the other models. 
\\

\par Overall, these findings suggest
that Random Forest and XGBoost are more adept at maintaining high performance with
fewer features, making them suitable choices for tasks where dimensionality reduction
is essential. KNN, while effective, may require additional tuning and careful feature
selection to achieve comparable performance, highlighting the importance of selecting
the appropriate model and dimensionality reduction technique based on the specific
classification task.

\section{Conclusion}
This study explores the predictive performance of three machine learning
models—Random Forest, k-Nearest Neighbours (KNN), and Extreme Gradient Boosting
(XGBoost)—for classifying solar flares into GOES categories. Using the dataset created by Liu et al. (2016), which includes 13 SHARP parameters, the models' effectiveness was compared for both binary and multiclass classification tasks. This comparison was conducted using 8 principal components (PC) capturing 95\% of the data variance and 100 PC capturing 97.5\% of the variance.
\par These findings indicate that Random Forest and XGBoost consistently
demonstrate strong performance across all metrics, benefiting significantly from the
increased dimensionality provided by 100 PC. This improved accuracy, ROC AUC, and
F1 scores underscore their ability to capture complex patterns and interactions in the
data. In contrast, while KNN shows modest improvement in multiclass classification
with additional principal components, its performance declines in binary classification,
suggesting a sensitivity to higher dimensionality and potential overfitting challenges.
\par These results emphasize the importance of dimensionality reduction and
careful model selection in solar flare prediction. Random Forest and XGBoost are
identified as robust models capable of maintaining high performance with fewer
features, making them suitable for tasks requiring dimensionality reduction. KNN,
though effective, may require additional tuning and feature selection to achieve
comparable performance.
\par The insights gained from this study can significantly enhance future scientific research
by improving solar flare prediction models, optimizing dimensionality reduction
techniques, and informing model selection for astrophysical tasks. The demonstrated
effectiveness of Random Forest and XGBoost in maintaining good performance with
fewer features suggests their applicability in various classification tasks beyond solar
flare prediction.
\par 
For future work, a refined solar flare prediction model will be trained and deployed using the most frequently occurring hyperparameter combinations for each model. Additionally, validated models will be integrated with real-time solar observation data to enhance the accuracy of existing space weather forecasting systems. This integration aims to improve the reliability of solar flare predictions, contributing to more effective early-warning mechanisms for space weather events.

\par
\begin{center}
\large \textbf{Acknowledgments}
\end{center}

I would like to thank the Fondation Jeunes Scientifiques Luxembourg (FJSL) for their support and resources, which were essential in the development of this project. Their encouragement and resources have been invaluable in carrying out this research. I also wish to acknowledge M. Olivier Parisot from the Luxembourg Institute of Science and Technology (LIST) for his valuable advice and guidance throughout this project.
\newpage

\footnotesize
\centering
\nocite{*}
\bibstyle{plain}

\end{document}